# Quantum Coherence of Edge States in the Quantum Hall Effect
## – Topological Invariants and Edge State Mixing –

Yasuhiro Hatsugai

*Department of Applied Physics, University of Tokyo 7-3-1 Hongo Bunkyo-ku, Tokyo 113, Japan* *

## Abstract

Edge states in the integral quantum Hall effect on a lattice are reviewed from a topological point of view. For a system with edges which is realized inevitably in an experimental situation, the Hall conductance $\sigma_{xy}$ is given by a winding number of the edge state on a complex energy surface. A relation between two topological invariants (bulk and edge) is also clarified. In a macroscopic system, mixture of the edge states are exponentially small and negligible. Quantum Coherence between the two edge states gives the quatization of $\sigma_{xy}$. However, when the system is mesoscopic, the mixture between the edges states plays a physical role. We also focus on this point and show some numerical results.

### I. INTRODUCTION AND THE LATTICE MODEL

To have a high precision of the quantization of $\sigma_{xy}$, topology plays a crucial role in the Quantum Hall effect. Electrons in a magnetic field feel a path dependent phase described by magnetic translations. This geometrical phase is essential and topological consideration is of

---

*e-mail: hatsugai@coral.t.u-tokyo.ac.jp

*central importance in the understanding of the effect. [1,2]. Contribution to the quantization of bulk states are exactly canceled by the gauge invariance and only edge states can contribute. The edge states on different edges are macroscopically separated. However, the quantum coherence between the edge states is responsible for the quantization of $\sigma_{xy}$.*

In the Landau gauge, a tight-binding hamiltonian for electrons on a lattice in a uniform magnetic field is given by $H = -\sum_{m,n}(c_{m+1,n}^\dagger c_{m,n} + c_{m,n+1}^\dagger e^{i2\pi\phi m} c_{m,n} + H.c.)$, where $c_{m,n}$ is the annihilation operator for a lattice fermion at site $(m,n)$. We assume that the magnetic field per plaquette $\phi = p/q$ with mutually prime integers $p$ and $q$. We always assume that the system is periodic in $y$ direction. Then the problem is reduced to a one-dimensional one with a parameter $k_y \in [0, 2\pi)$ which lies in a magnetic Brillouin zone ($0 \equiv 2\pi$). The one-dimensional Schrödinger equation $H(k_y)|\Psi(k_y)\rangle = E(k_y)|\Psi(k_y)\rangle$ is given as

$$H(k_y)|\Psi(k_y)\rangle \equiv -\{\Psi_{m+1}(k_y) + \Psi_{m-1}(k_y)\} - 2\cos(k_y - 2\pi\phi m)\Psi_m(k_y) = E(k_y)\Psi_m(k_y). \tag{1.1}$$

## II. THE CHERN NUMBER: $\sigma_{XY}$ OF A BULK SYSTEM

If the system is infinite, that is, without edges [1], above one-dimensional hamiltonian has a period $q$. Therefore the Bloch theorem tells that the Bloch function $\Psi_m$ satisfies $\Psi_m(k_x, k_y) = e^{ik_x m}\tilde{u}_m(k_x, k_y)$ with $\tilde{u}_{m+q}(\mathbf{k}) = \tilde{u}_m(\mathbf{k})$ and $k_x \in [0, 2\pi/q)$, $0 \equiv 2\pi/q$ where $\mathbf{k}$ is defined on a magnetic Brillouin zone which is a two-dimensional torus. The spectrum consists of $q$ energy bands and $\sigma_{xy}$ of the filled $j$-th band ($j = 1, \cdots, q$) is calculated by the Kubo formula as [1]

$$\sigma_{xy}^{j,\,bulk} = -\frac{e^2}{h}\frac{1}{2\pi i}\int\int_{T_{MBZ}^2} dk_x dk_y\,[\,\nabla_k \times \mathbf{A}_u^j(\mathbf{k})\,]_z, \tag{2.1}$$

where $\mathbf{A}_u^j(\mathbf{k}) = \langle u^j(\mathbf{k})|\nabla_k|u^j(\mathbf{k})\rangle = \sum_{i=1}^q u_m^{j\,*}(\mathbf{k})\,\nabla_k u_m^j(\mathbf{k})$ and $u_m^j$ is obtained from $\tilde{u}_m^j$ by normalizing as $\langle u^j(\mathbf{k})|u^j(\mathbf{k})\rangle = 1$.

In this form, the phase of the Bloch wave function defines $U(1)$ vortices on the magnetic Brillouin zone and the total vorticity which is the Chern number gives the Hall conductance.

This is a topological meaning of the Hall conductance of the bulk states.

## III. THE WINDING NUMBER: $\sigma_{XY}$ OF A CYLINDRICAL SYSTEM

Now let us consider a system in a cylindrical geometry, *i.e.*, the system is periodic in y direction and finite in x direction. Here we only state the results of Ref. [2] relevant to our purpose. The boundary condition for the cylindrical geometry is

$$\Psi_0^{(e)}(k_y) = \Psi_{L_x}^{(e)}(k_y) = 0 : \text{with edges} \tag{3.1}$$

Mathematically the difference between systems with and without edges is a difference in the boundary conditions. For the system without edges, the condition follows the Bloch theorem as

$$\Psi_{m+q}^{(b)}(k_y) = \rho \Psi_m^{(b)}(k_y), \ \rho = e^{iqk_x} : \text{without edges} \tag{3.2}$$

The strategy of the work is roughly as follow. *The edge states are a sort of bound states and the Bloch states are scattering states. These two solutions are treated in a unified way when one analytically continues the energy to a complex energy. A number of the bound states are governed by an information of the scattering states which is known in the continuum scattering theory.*

The Schrödinger equation Eq. (1.1) is written as

$$\begin{pmatrix} \Psi_{m+1} \\ \Psi_m \end{pmatrix} = \tilde{M}_m(E, k_y) \begin{pmatrix} \Psi_m \\ \Psi_{m-1} \end{pmatrix}, \ \tilde{M}_m(E, k_y) = \begin{pmatrix} -E - 2\cos(k_y - 2\pi\phi m) & -1 \\ 1 & 0 \end{pmatrix}. \tag{3.3}$$

All solutions are obtained by different choices of an initial condition $\Psi_0$ and $\Psi_1 = 1$ fixes the normalization. For the cylindrical geometry, we use $\Psi_0^{(e)} = 0$ and $\Psi_1^{(e)} = 1$. Then the roots of the algebraic equation $\Psi_{L_x}^{(e)} = [M(\epsilon)]_{21}^l = 0$ gives the energies of the system where $M(E) = \tilde{M}(E)_q \tilde{M}(E)_{q-1} \cdots \tilde{M}(E)_2 \tilde{M}(E)_1$.

There are $q$ energy bands and each energy gap has one edge state, *i.e.*, there are $g = q - 1$ edge states. These energies of the edge state $\mu_j$ ($j = 1, \cdots, g$) are determined by a $q$ site problem as [2]

$$\Psi_q^{(e)}(\mu_j) = M_{21}(\mu_j) = 0 : \text{with edges}. \quad (3.4)$$

On the other hand, the Bloch function satisfies Eq. (3.2). Thus $\Psi_1$ and $\Psi_0$ form an eigenvector of $M$ with the eigenvalue $\rho$ as

$$M(E)\begin{pmatrix} \Psi_1^{(b)} \\ \Psi_0^{(b)} \end{pmatrix} = \begin{pmatrix} \Psi_{q+1}^{(b)}(E) \\ \Psi_q^{(b)}(E) \end{pmatrix} = \rho(E)\begin{pmatrix} \Psi_1^{(b)} \\ \Psi_0^{(b)} \end{pmatrix} : \text{without edges}. \quad (3.5)$$

We analytically continue $E$ to a complex energy $z$ in order to construct a wave function of the edge state from that of the Bloch function.

Here

$$\rho(z) = \frac{1}{2}(\Delta(z) - \sqrt{\Delta(z)^2 - 4}) = e^{iqk_x} \quad (3.6)$$

with $\Delta(z) = \text{Tr } M(z)$. For a bulk state, $k_x$ is real but is imaginary for the edge states. The analytic structure of the wave function is determined by the function $\omega = \sqrt{\Delta(z)^2 - 4}$. A Riemann surface of the $\omega^2 = \Delta(z)^2 - 4$ which is constructed by gluing two complex energy surfaces ( Riemann spheres, $R^+$ and $R^-$ ). Each of them has branch cuts given by $\Delta(z)^2 - 4 \leq 0$ at $\Im z = 0$. Since this condition gives $|\rho| = 1$, the branch cuts are given by the $q$ energy bands. Thus the Riemann surface is obtained by gluing $R^+$ and $R^-$ at these $q$ energy bands. (Fig . 1). The genus of the surface is $g = q - 1$ which is the number of energy gaps. The complex energy surface of the one dimensional system Eq. (1.1) is a genus $g = q - 1$ Riemann surface $\Sigma_g(k_y)$ for each $k_y$.

On $\Sigma_g(k_y)$, the energy gaps correspond to circles around the holes of the $\Sigma_g(k_y)$ and the energy bands correspond to closed paths on $\Sigma_g(k_y)$. $\Psi_q^{(b)}$ has always $g$ zeros at $\mu_j$ ( $\Psi_q^{(b)}(\mu_j) = 0$) which give the edge state energies. Changing $k_y$ from 0 to $2\pi$, we obtain a family of the Riemann surface $\Sigma_g(k_y)$ which is topologically equivalent if there are stable energy gaps in the two dimensional spectrum. By identifying the $\Sigma_g(k_y)$, the $\mu_j(k_y)$ moves around the holes and forms an oriented loop $C(\mu_j)$. This winding number of the loop around the $j$-th hole $I(C(\mu_j))$ is a well defined topological quantity which gives $\sigma_{xy}$. When the Fermi energy of the two dimensional system lies in the $j$-th energy gap, $\sigma_{xy}$ is given by this winding number as [2]

$$\sigma_{xy}^{edge} = \sum_{l=1}^{j} \sigma_{xy}^{l} = -\frac{e^2}{h} I(C(\mu_l)). \qquad (3.7)$$

This is the second interpretation of the Hall conductance as a topological number.

Further, one can see that the zeros of $\Psi_q^{(b)}(k_x, k_y)$ on the energy bands are given by the points where the edge state is degenerate with the bulk state. Counting a sum of the vorticities on the energy band which come from the degenerate edge states, the Chern number of the $j$-th band is written by the winding number as $I(C_j) - I(C_j - 1)$. Therefore the Hall conductance for the filled $j$-th band is

$$\sigma_{xy}^{j,\ bulk} = -\frac{e^2}{h}[\ I(C_j) - I(C_{j-1})\ ] = \sigma_{xy}^{j,\ edge}. \qquad (3.8)$$

This is the relation between the Chern number for the bulk state and the winding number of the edge state. It clearly demonstrates the relation between the two interpretations of the Hall conductance with and without edges.

## IV. EDGE STATE MIXING IN A MESOSCOPIC SYSTEM

When the system is macroscopic, edge states on each edges of the system are separated macroscopically. Thus the mixture of the edge states are exponentially small and negligible. This is true even in a dirty sample. This is a typical realization of the "Quantum Coherence" and it gives the high accuracy of the quantization. The quantum coherence between macroscopically separated edge states gives the Quantum Hall effect. However, when the system is mesoscopic, the mixture between the edges states plays a physical role. We have obtained wavefunctions on a pinched cylinder with randomness numerically. It models a tunneling between two edge states. In Fig. 2, the spectral flow of the system is shown as a function of the Aharonov-Bohm flux through the cylinder $\phi$. If there is no mixture between the edge states, the spectral flow shows a level crossing, which gives non zero Hall conductance. When there is a mixture, the level repels each other as shown in the Figure. The corresponding wave functions are shown in Figs. 3 and it clearly show the mixture of the edge states. This is only possible in a mesoscopic system. One important and interesting fact is that the behavior of the edge state is very sensitive to the AB flux. The localized bulk states are almost insensitive but the edge states are not. Changing the AB flux just a fraction of the flux quantum, one can move the edge state from one edge to the other macroscopicaly (mesoscopically). It may suggest a possibility for device applications.

FIGURES

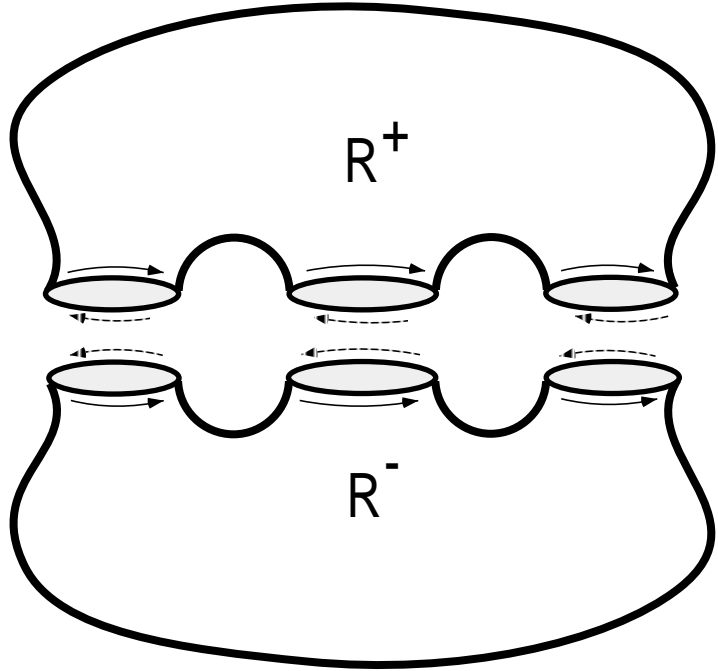

FIG. 1. The complex energy surface which is composed of $R^+$ and $R^-$. The genus (number of holes) is a number of the gaps.

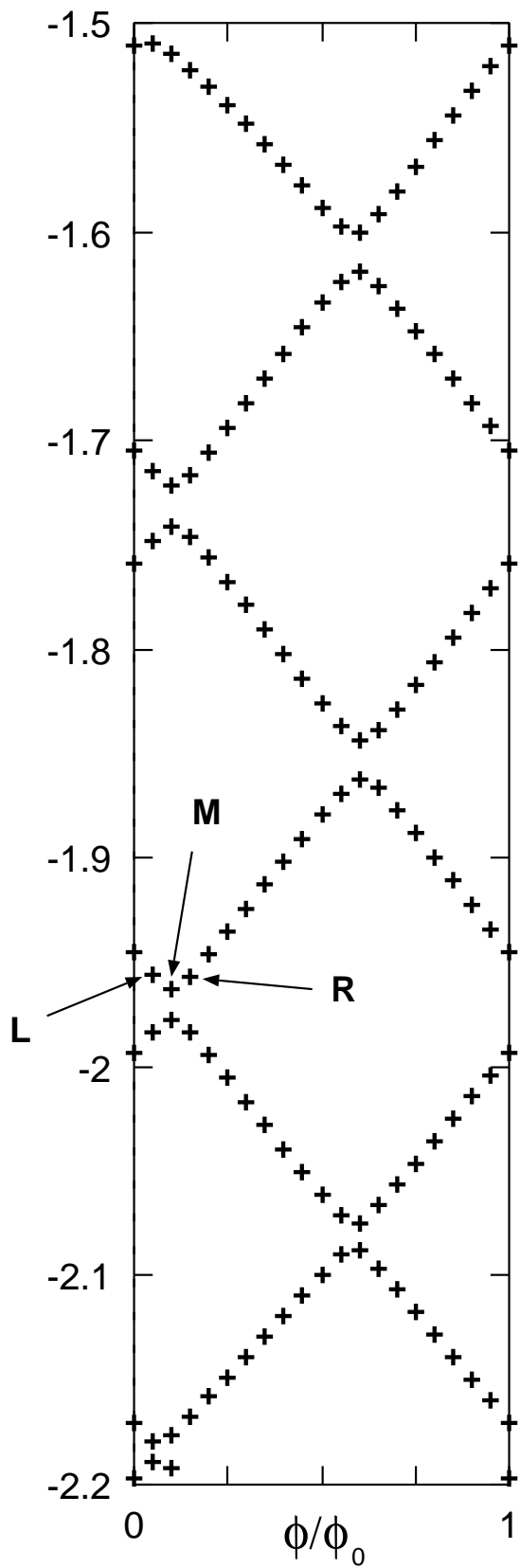

FIG. 2. Spectral flow of the edge states as a function of the Aharonov Bohm flux $\phi/\phi_0$.

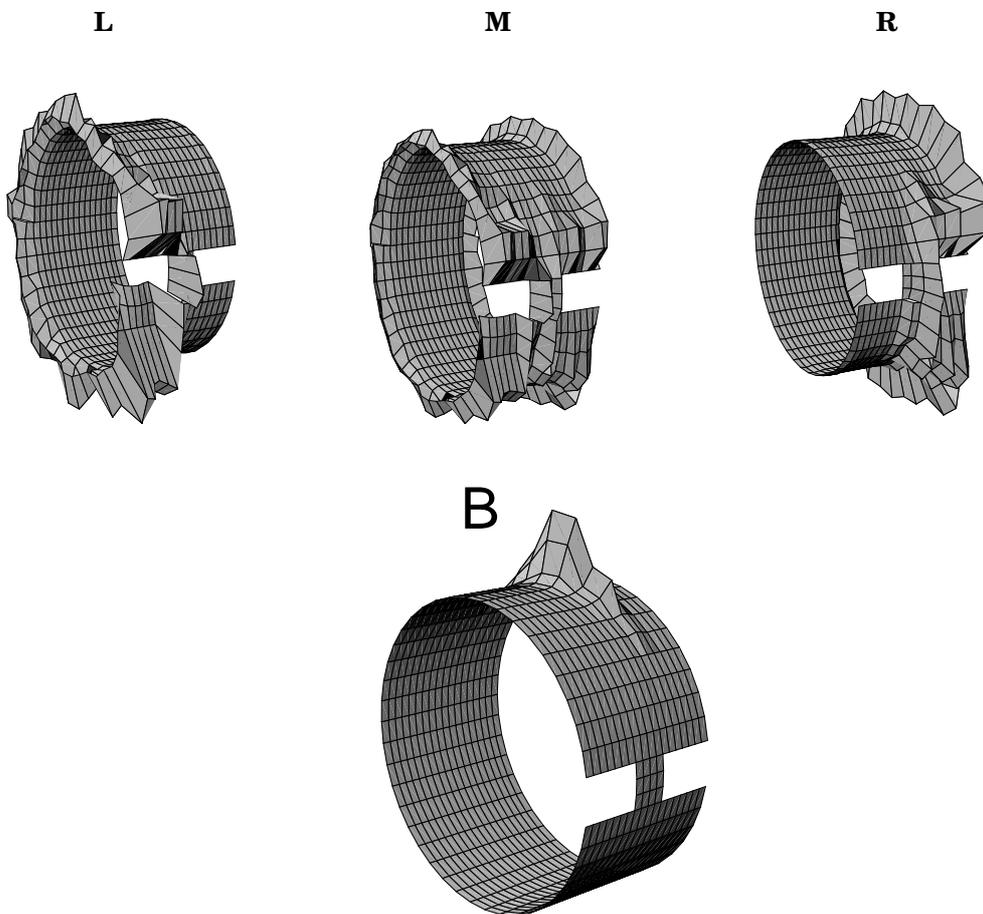

FIG. 3. Wave functions of the edge states on a pinched cylinder. L, M, R: edge states, B: a bulk state.

# REFERENCES

[1] D. J. Thouless, M. Kohmoto, P. Nightingale, and M. den Nijs, Phys. Rev. Lett. **49**, (1982) 405; M. Kohmoto, Ann. Phys. (N. Y. ) **160**, (1985) 355.

[2] Y. Hatsugai, Phys. Rev. B **48**, (1993) 11851; Phys. Rev. Lett. **71**, (1993) 3697.